\documentclass[]{aa}
\usepackage{graphicx,epsf,psfig,epsfig}
\newcommand{\un}[1]{~\hspace{-2pt}\ensuremath{\mathrm{#1}}}
\usepackage{color}
\begin{document}
   \title{First Results from the IBIS/ISGRI Data Obtained During the Galactic Plane Scan\thanks{This paper 
is based on observations with INTEGRAL, an ESA project 
with instruments and science data centre funded by ESA member states 
(especially the PI countries: Denmark, France, Germany, Italy, Switzerland, 
Spain), Czech Republic and Poland, and with the participation of Russia 
and the USA.}}

   \subtitle{II. The Vela Region}

   \author{J. Rodriguez\inst{1,2}, M. Del Santo\inst{3}, F. Lebrun\inst{4}, G. Belanger\inst{4}, M. Cadolle-Bel\inst{4}
F. Capitanio\inst{3}, P. David\inst{1}, L. Foschini\inst{5}, P. Goldoni\inst{4}, 
A. Goldwurm\inst{4}, A. Gros\inst{4}, P. Laurent\inst{4}, A. Paizis\inst{6,2}, J. Paul\inst{4}
R. Terrier\inst{1}, S. E. Shaw\inst{7,2}, P. Ubertini\inst{3}
}
   \offprints{J. Rodriguez : jrodriguez@cea.fr}

   \institute{CEA Saclay, DSM/DAPNIA/SAp (CNRS FRE 2591), F-91191 Gif sur Yvette Cedex, France 
          \and  Integral Science Data Center, Chemin d'Ecogia, 16, CH-1290 Versoix
Switzerland
\and IASF/CNR, via del Fosso del Cavaliere 100, 00133 Roma, Italy
\and CEA Saclay, DSM/DAPNIA/SAp, F-91191 Gif sur Yvette Cedex, France
\and IASF/CNR, sezione di Bologna, via Gobetti 101, 40129 Bologna, Italy
\and IASF/CNR, Sezione di Milano, Via Bassini 15, 20133 Milano, Italy
\and School of Physics and Astronomy, University of Southampton, Southampton, SO17 1BJ UK.
            }
\authorrunning{J. Rodriguez et al.}
\titlerunning{Galactic Plane Scan: Vela Region}

   \date{Received  ; accepted}

   \abstract{We report on INTEGRAL/IBIS observations of the Vela region
during a Galactic Plane Scan (hereafter GPS) presenting the IBIS in-flight
performances during these operations. Among all the known sources in the field
 of view we clearly detect 
 4U~0836$-$429, Vela~X$-$1, Cen~X$-$3, GX~301$-$2, 1E~1145.1$-$6141, 
and H0918$-$549 in the 20-40 keV energy range. Only Vela X$-$1 and GX~301$-$2
are detected in the 40$-$80 keV energy range, and no sources are visible above.
We present the results of each individual observation
($\sim 2200$ s exposure), as well as those from the mosaic of these scans. 
   }

   \maketitle

\section{Introduction}
The IBIS telescope (Ubertini et al. 2003) on board the INTEGRAL observatory
is a $\gamma$-ray sensitive 
coded mask telescope. It is composed of two layers, each one being a detector. 
The upper one, ISGRI (Lebrun et al. 2003), is sensitive 
between 15\un{keV} and 1\un{MeV}, and is optimized between 15\un{keV} 
and $\sim 200$\un{keV}. The lower one, PICSIT (Di Cocco et al. 2003),
 is sensitive between $\sim 175$\un{keV} and $\sim 10$\un{MeV}. Due to 
its high angular resolution ($12^{\prime}$ FWHM), and wide field of view 
(FOV, 19$^{\circ}\times$19$^{\circ}$ at half response) IBIS is 
perfectly suited for the discovery and localisation of new transient X-ray 
and $\gamma$-ray sources.  The energy range and the high sensitivity of 
ISGRI are particularly well adapted to perform a hard X-ray survey of the
Galactic plane. In order to exploit these features $\sim 8\%$ of 
the observing time is 
devoted to scans of the Galactic plane (GPS) during the first year of the nominal
 mission. For a precise description of the GPS see Winkler et al. (2003).\\
\indent Such scans 
are particularly important to follow up transient and persistent 
X-ray sources, and for the discovery of new X-ray nov\ae. Up to now, 
more than 10 hard X-ray sources have been  (re-)discovered mainly with the
ISGRI layer of the IBIS detector. 
The INTEGRAL observations of the first such source, IGR J16318$-$4848 
(Courvoisier et al. 2003), are presented by Walter et al. (2003a). 
Most of these sources belong to the class of X-ray binaries (XB),
 accretion powered binary systems that host a compact object (neutron star 
or black hole), and that can be persistent or transient. X-ray 
transients (XT) spend most of their 
life in quiescence, and are detected in the X-rays as they undergo episodes 
of outburst. These outbursts can be  
 recurrent  (e.g. 4U1630$-$47, XTE J1550$-$564) or only detected 
once (Nova Musca 1991). Persistent sources are always visible in
X and $\gamma$-rays (e.g. Cyg X$-$1, or GX 339$-$4). In the former case,  
ISGRI's sensitivity allows the detection of an outburst well before 
the other all-sky monitors, especially
for outbursts starting with periods of hard states (e.g. XTE J1550$-$564, 
Dubath et al. 2003), but also for investigating the late stages of outbursts,
 that are poorly studied in this spectral 
range up to now (e.g. XTE J1720$-$318, Goldoni et al. 2003).
The IBIS spectral range is also particularly well suited to discover, or 
re-discover
heavily absorbed sources, which were missed during previous 
observations in the soft 
(1$-$10\un{keV}) band, where the X-rays are very sensitive to the absorption 
along the line of sight.
The high localisation accuracy of INTEGRAL's imaging software
 ($\sim 1^{\prime}$, Gros et al. 2003) and the understanding of the alignment
systematics (Walter et al. 2003b) have already 
allowed for three multiwavelength counterpart searches and/or follow-up 
observations in softer X-rays with high sensitivity X-ray telescopes such 
as XMM-Newton (IGR J16318$-$4848, Matt \& Guainazzi 2003; IGR J16320$-$4751,  
Rodriguez et al. 2003), and Chandra (IGR J16358$-$4726 Kouveliotou et al. 2003).
 Such multiwavelength observations are the best tool for identifying the 
nature of a source.\\
\indent Other types of sources are also known to produce significant X-ray and 
$\gamma$-ray emission:
anomalous X-ray pulsars, isolated pulsars, AGN seen through the Galactic 
plane, and high energy sources such as those detected by the EGRET telescope. 
The unprecedented localisation accuracy of IBIS over its energy domain
will without a doubt help in identifying some of the
so-called "unidentified" EGRET sources, as well as their possible
X- and $\gamma$-ray counterparts several of which are often found
within EGRET's large error box.\\
\indent After a first scan in the Cygnus region (Del Santo et al. 2003, 
hereafter paper 1), where the telescope
has demonstrated its ability in imaging bright sources like Cyg X$-$1, a 
second scan was performed on the other side
of the Galactic plane, starting on the Vela region and finishing around 
the Centaurus region. We present here the results obtained during these 
observations, where many fainter sources are detected. Technical aspects of 
the observations and data reduction method
are presented in the following section. The results are presented
in section~3 and discussed in section~4.

\section{Observations and Data Reduction}
\begin{figure}[htbp]
\epsfig{file=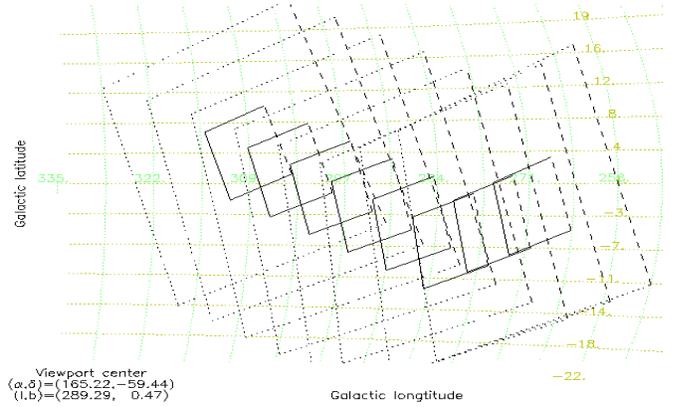,width=\columnwidth,height=5.5cm}
\caption{Scheme of the GPS performed during Rev. 30. The scan started at 
l=270.0 $^\circ$ and finished at l=309.5 $^\circ$. The continuous squares represent the limits of the fully coded field of view, and the dashed squares the limit of the partially coded field of view.}
\label{fig:map}
\end{figure}

The GPS considered here was carried out on January 11$^{\mathrm{th}}$ 
2003, it is composed of 8 $\times$ 2200\un{s} pointings. The scan started at 
11h51 UTC, and finished at 17h35 UTC. The first individual observation 
was centered on 
l=270.0$^\circ$, and the last one on l=309.5$^\circ$ (Fig. \ref{fig:map}).
The log of the observations is shown in Table \ref{tab:log}.

\begin{table}[htbp]
\caption{Log of the observations performed during the scan.}
\begin{tabular}{c c c c}
Obs. Number &  time      & \multicolumn{2}{c}{Pointing direction}\\
            &   (UTC)    &   R.A.       & Dec \\
\hline
\hline
1           & 11h51      & $09^{\mathrm{h}}$ $02^{\mathrm{m}}$ $27.22^{\mathrm{s}}$ & $-49^\circ$ $47'$ $52.7''$\\
2           & 12h57      & $09^{\mathrm{h}}$ $17^{\mathrm{m}}$ $12.99^{\mathrm{s}}$ & $-55^\circ$ $20'$ $58.0''$\\
3           & 13h37      & $09^{\mathrm{h}}$ $37^{\mathrm{m}}$ $06.03^{\mathrm{s}}$ & $-60^\circ$ $47'$ $39.5''$\\
4           & 14h17      & $10^{\mathrm{h}}$ $25^{\mathrm{m}}$ $34.76^{\mathrm{s}}$ & $-62^\circ$ $32'$ $12.4''$\\
5           & 14h58      & $11^{\mathrm{h}}$ $18^{\mathrm{m}}$ $36.64^{\mathrm{s}}$ & $-63^\circ$ $09'$ $17.5''$\\
6           & 15h38      & $12^{\mathrm{h}}$ $10^{\mathrm{m}}$ $57.27^{\mathrm{s}}$ & $-62^\circ$ $30'$ $19.5''$\\
7           & 16h18      & $12^{\mathrm{h}}$ $59^{\mathrm{m}}$ $12.47^{\mathrm{s}}$ & $-60^\circ$ $42'$ $38.2''$\\
8           & 16h58      & $13^{\mathrm{h}}$ $41^{\mathrm{m}}$ $07.60^{\mathrm{s}}$ & $-57^\circ$ $57'$ $48.4''$\\
\hline
\end{tabular}
\label{tab:log}
\end{table}
\indent The images were generated with the standard pipeline developed 
in cooperation by CEA/Saclay and ISDC/Versoix (see e.g Goldwurm et al. 2001, 2003).
For each pointing we have produced images in 3 energy ranges, 20$-$40\un{keV}, 
40$-$80\un{keV}, 80$-$160\un{keV}.  Note that PICsIT works in a higher 
energy range (above 200\un{keV}), 
where very long exposures ($10^5-10^6$\un{s}) are required to obtain 
significant detections. PICsIT data are not considered further in this analysis.
All the pointings were then combined to produce a mosaic. The image analysis 
is performed in a way such that the positions
of all detected excesses are compared to catalog (Ebisawa et al.
 2003) source positions and then fitted to obtain a fine position (Goldwurm 
et al. 2001,  2003).
Any excess whose position does not correspond to a catalog source
is labeled as a new source. No such new source was found in
our analysis.
Only sources with a significance level greater than 5 and for
which the fitting procedure converged are considered 
in this paper.

\section{Results}
\begin{figure*}[htbp]
\centering
\begin{tabular}{cc}
\epsfig{file=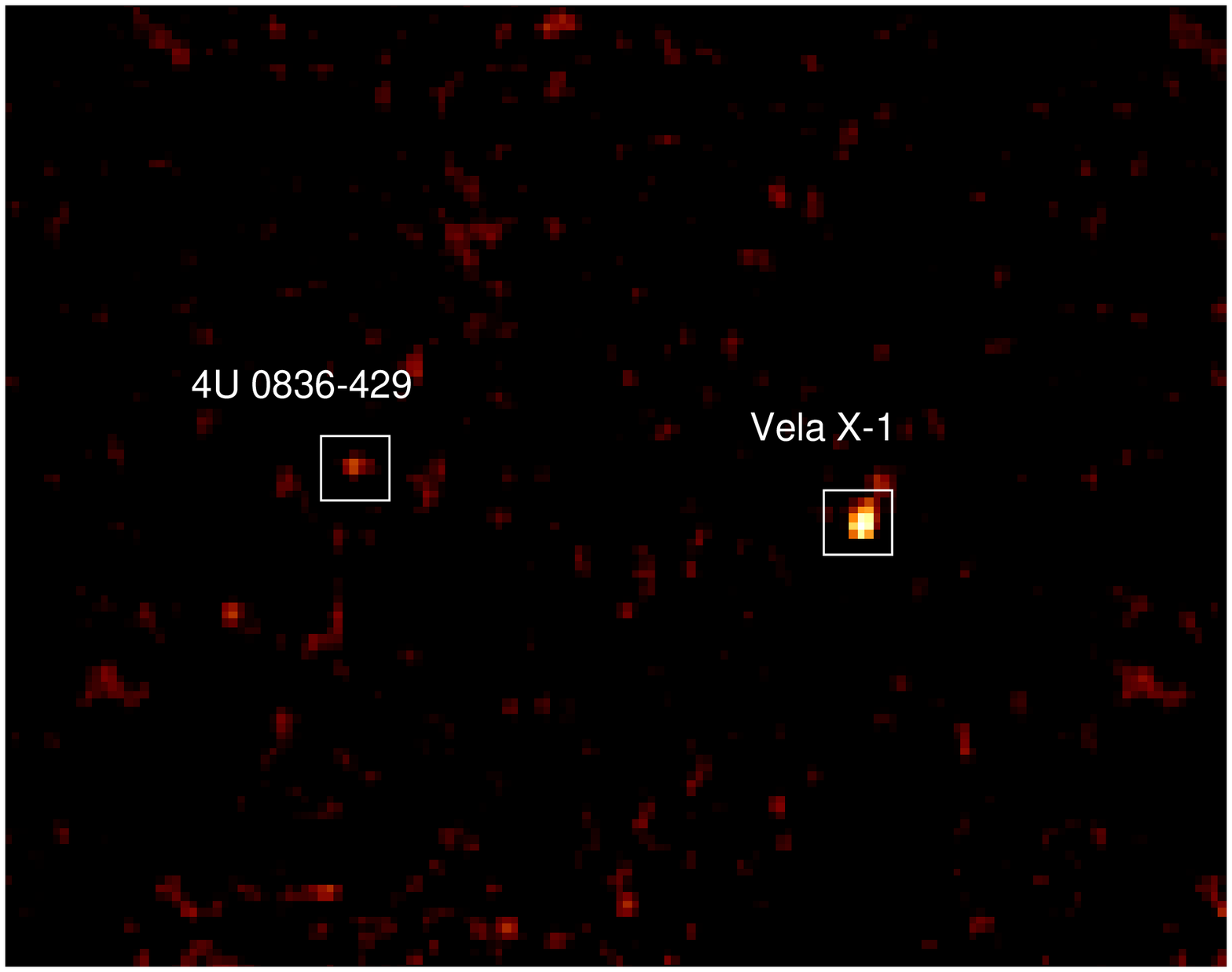,width=6cm,height=4cm}&
\epsfig{file=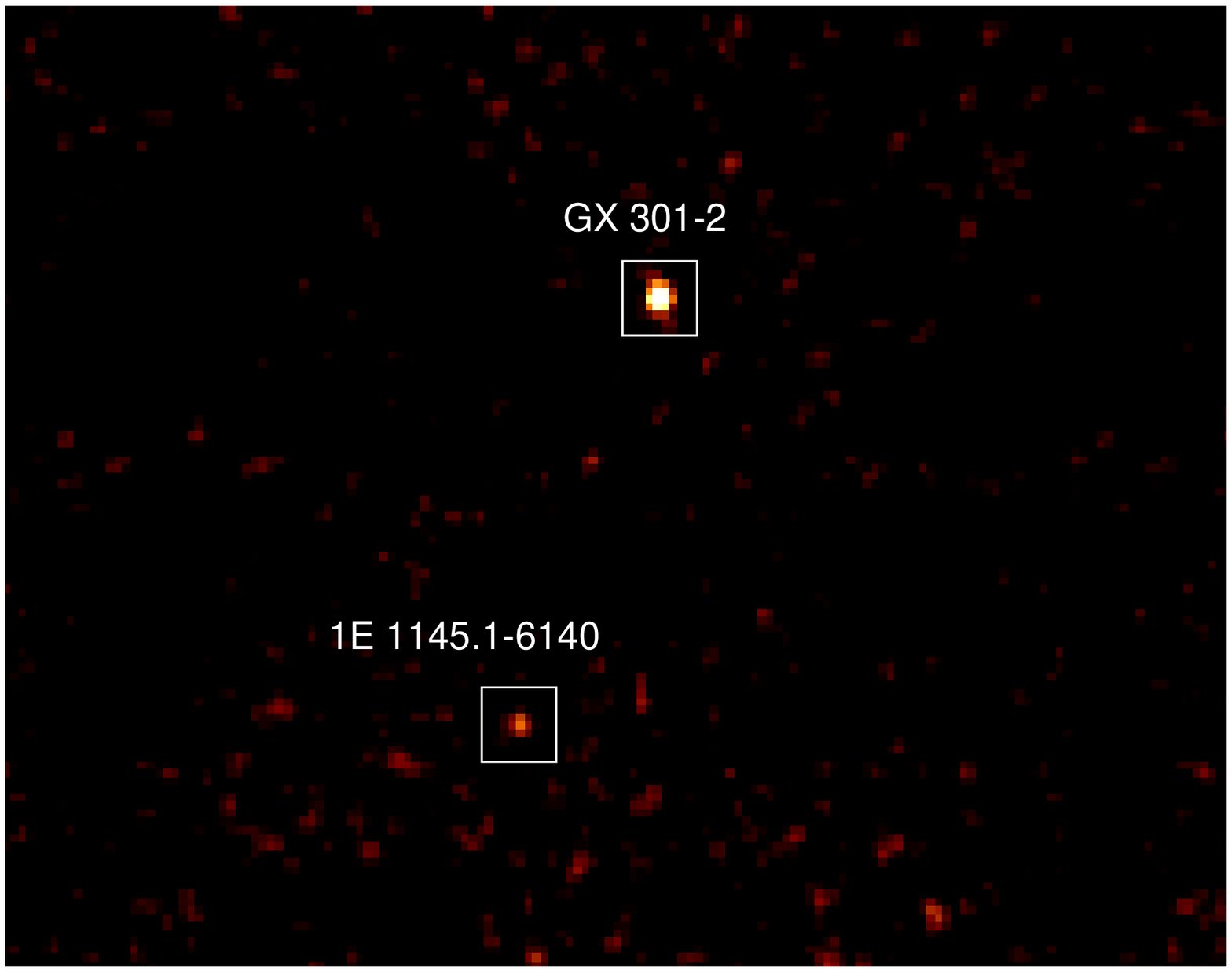,width=6cm,height=4cm}\\
\end{tabular}
\caption{12.6$\times$10$^{\circ}$ zoom on the individual 20-40\un{keV} image 
from obs \#1 , and \#6. 
This sequence illustrates
the sensitivity achieved in individual pointings. GX 301$-$2, Vela X$-$1, 
4U 0836$-$429, 1E 1145.1$-$6141 are 
clearly detected by the software. 4U 0836$-$429, is not detected in 
the mosaic. 
Note that there are still some residuals/background structures in the images. The precise background correction is discussed in Terrier et al. (2003).}
\label{fig:indiv}
\end{figure*}

\begin{figure*}[htbp]
\centering
\epsfig{file=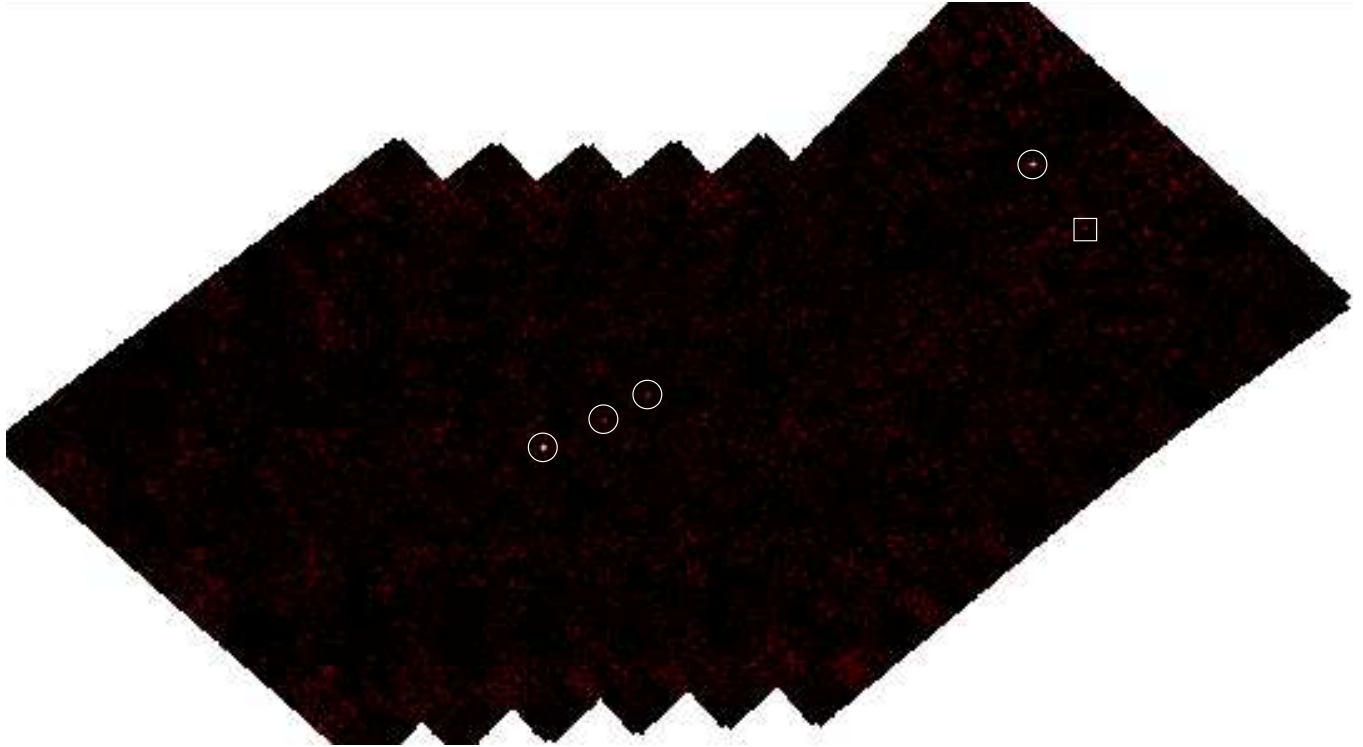,width=18cm, height=10cm}
\caption{20$-$40\un{keV} mosaic image produced by summing each individual image. 
Four sources are detected by the software (circles), from right to left these 
are  Vela X-1 Cen X-3, 1E 1145.1-6141, and GX 301$-$2.  
Although not detected 4U0836-429 (square) is visible on the image.}
\label{fig:mosa}
\end{figure*}

The 20$-$40\un{keV} images obtained for some individual pointings are 
presented in Fig. \ref{fig:indiv},
and the 20$-$40\un{keV} mosaic produced from the whole scan on Fig. \ref{fig:mosa}.

The scientific parameters such as the source position  and the flux 
returned from the software are 
reported in Table \ref{tab:offset}. For each of them we estimate 
the offset between the position found with ISGRI and the catalogue 
(based on the SIMBAD database) position.
 The angle from the center of the FOV is also reported. It is clear that a 
precise analysis of a given GPS requires 
that each pointing is precisely studied along with the mosaic resulting from the summation
 of all pointings.
Variable sources can appear in unique pointings, and be 
absent in the following ones, and therefore not be detected in a mosaic  because the S/N is less
than for the pointing where the source is bright. This is probably the case for H0918$-$549 
and 4U 0836$-$429, both variable sources. Note
 that 4U 0836$-$429 can only be seen in two pointings, and that it is at the edge
of the FOV (where the sensitivity is much lower) in the second one, which can also explain 
why it is detected only in the first pointing.
On the other hand, steady faint sources will 
not appear in a single pointing, but need more accumulation time to 
be detected.

\begin{table}[htbp]
\caption{Scientific parameters returned by the software; position (after having applied the 
misalignment matrix, Walter et al 2003b) and flux of 
the detected sources in the 20$-$40\un{keV} energy range, for each individual 
pointing, and the final mosaic. 
We have put in addition the offset between the estimated position,
 and  the real position as given by the SIMBAD online archives. The off axis angle is also given for each pointing.
The source types are extracted from the catalogues of Liu et al. (2000, 2001).}
\begin{tabular}{c c c c c }
Source & Type  & Flux  & offset & offaxis \\
       &      &(mCrab)& (arcmin) & angle\\
\hline
\hline
\multicolumn{5}{c}{Pointing n$^\circ$ 1}\\
\hline
Vela X$-$1 & HMXB & $79\pm 5$ & 2.3 & 9.2$^{\circ}$   \\
4U 0836$-$429 & LMXB & $38\pm 6$ & 2.7 & 8.0$^{\circ}$\\
\hline
\multicolumn{5}{c}{Pointing n$^{\circ}$ 2}\\
\hline
H0918$-$549 & LMXB & 21$\pm4$ & 1.8 & 29 arcmin\\
\hline
\multicolumn{5}{c}{Pointing n$^{\circ}$ 4}\\
\hline
Cen X$-$3  & HMXB & 32$\pm5$ & 2.7 &   6.9$^\circ$   \\
GX 301$-$2 & HMXB & 59$\pm9$ & 2.9 &   13.8$^\circ$   \\
\hline
\multicolumn{5}{c}{Pointing n$^{\circ}$ 5}\\
\hline
Cen X$-$3  & HMXB & 32$\pm4$ & 1.0 &   2.5$^\circ$   \\
GX 301$-$2 & HMXB & 53$\pm5$ & 0.6 &   7.7$^\circ$   \\
\hline
\multicolumn{5}{c}{Pointing n$^{\circ}$ 6}\\
\hline
GX 301$-$2 & HMXB & 75$\pm4$ & 0.7 &   1.8$^\circ$   \\
1145.1$-$6141 & HMXB & 30$\pm5$ &  1.6 & 2.8$^\circ$   \\
\hline
\multicolumn{5}{c}{Pointing n$^{\circ}$ 7}\\
\hline
GX 301$-$2 & HMXB & 76$\pm5$ & 1.1 &   4.3$^\circ$   \\
1145.1$-$6141 & HMXB & 22$\pm4$ &  0.2 & 8.6$^\circ$   \\
\hline
\multicolumn{5}{c}{Pointing n$^{\circ}$ 8}\\
\hline
GX 301$-$2 & HMXB & 70$\pm5$ & 3.5 &   10.3$^\circ$   \\
\hline
\multicolumn{5}{c}{Result of the mosaic}\\
\hline
\multicolumn{2}{c}{Source Name} &\multicolumn{2}{c}{Flux}& offset\\
& & \multicolumn{2}{c}{mCrab}& arcmin\\
\hline
\multicolumn{2}{c}{Vela X$-$1} &\multicolumn{2}{c}{91$\pm$7} & 2.4\\
\multicolumn{2}{c}{GX 301$-$2} &\multicolumn{2}{c}{71$\pm$4} & 0.8\\
\multicolumn{2}{c}{Cen X$-$3}  &\multicolumn{2}{c}{27$\pm$3}& 2.2\\
\multicolumn{2}{c}{1145.1$-$6141}  &\multicolumn{2}{c}{20$\pm$3}& 2.0\\
\hline
\hline
\end{tabular}
\label{tab:offset}
\end{table}

\section{Discussion}
From the analysis of each individual pointing,
sources with a relatively low level of flux ($\sim 20$ mCrab in the 
20$-$40\un{keV} energy range) have been detected with a relatively good
 accuracy 
on the estimated position (1.8 $^{\prime}$ offset for H0918$-$549 
Table \ref{tab:offset}). This level of flux 
seems to be the  sensitivity limit of ISGRI at the current 
stage of development\footnote{$\sim 5\sigma$ detection level for an exposure 
of 2200s} (see also paper 1). Note that for large angles from the 
center of the FOV, the 
flux of a source is still rather uncertain at this early stage of the mission. 
Fainter sources can be detected with longer exposure times, as e.g. 
1E1145.1$-$6141, which, has 
an averaged flux of about 20 mCrab, over the four pointings where it is in the
FOV.  Some slight offset (up to 3.5 $^{\prime}$) 
is still found between the estimated positions and the 
catalogue ones.
It is worth noting that the offset depends on 
two factors: the distance from the center of the FOV, and the source flux 
(thus the significance of its detection). As the source is far from the centre,
the offset increases (Table \ref{tab:offset}), while for bright sources 
(sources detected with higher significance) the location accuracy usually 
increases  (see Gros et al. 2003, Walter et al. 2003b and paper 1). In the 40$-$80 keV 
energy range
only Vela X$-$1 and GX 301$-$4 are detected, and no source is seen above.\\
\indent One important aspect of such scans is that they allow 
source evolution  to be followed on long time scales. As mentioned in the 
introduction, they enable transient source to be detected as they undergo 
outbursts. However, as illustrated here in the particular case
of GX 301$-$2, and thanks to the large FOV of the IBIS detector, such 
scans are particularly important to construct light curves of persistent 
sources, during the time they are seen in the FOV of the
telescope. By repeating those GPS patterns from time to time, it is possible
 to monitor the long term behaviour of those sources.\\
\indent The ISGRI camera of the IBIS detector onboard INTEGRAL has largely
proven its utility in such scans, after only $\sim 6$ months of activity. 
Many of the new sources found up to now were detected during scans of the 
Galactic plane or of the Galactic Center region. With its high positioning 
accuracy it allowed for the identification of counterparts, and for 
X-ray follow-up (Revnivtsev et al 2003, Rodriguez et al. 2003). It appears 
thus as an instrument of primary importance in 
multiwavelength campaigns dedicated to the study of sources of 
high energy, which are the only valuable way to understand properly the 
physical mechanisms in action in those sources. 
     
\begin{acknowledgements}
The authors warmly thank T. Courvoisier for careful reading and useful 
comments who helped to improved the quality of the paper.
J.R. acknowledges financial support from the French Spatial Agency (CNES).
M.D.S. and L.F. acknowledge financial support from Italian Space Agency (ASI), and the hospitality of the ISDC.
 \end{acknowledgements}

\end{document}